\documentclass[aps,prc,floatfix,groupedaddress,showpacs,amsfonts,nofootinbib,twocolumn]{revtex4}
\usepackage{bm}
\usepackage{epsfig}
\usepackage{amsmath}
\usepackage{dcolumn}

\begin{document}

\title{Relativity versus exchange currents in $^{16}$O($e,e^{\prime}p$)}

\author{J. Grineviciute}
\affiliation{Faculty of Physics, Warsaw University of Technology, ulica Koszykowa 75 Warsaw, Poland}

\author{Dean Halderson}
\affiliation{Department of Physics, Western Michigan University, Kalamazoo, MI 49008}

\begin{abstract}
\begin{description}
\item[Background:] The $^{16}$O($e,e^{\prime}p$) reaction in the quasielastic region has been studied in several experiments to determine spectroscopic factors, hence, the degree to which $^{16}$O looks like a closed shell.  By varying the kinematics, experimentalists are able to extract response functions which comprise the cross section.  However, analysis of the response functions separately produces very different spectroscopic factors. Two calculations led to different conclusions as to whether exchange currents can eliminate the discrepancies.  Neither calculation considered relativistic corrections.

\item[Purpose:] The purpose of the article is to investigate the disagreement as to whether exchange currents are the solution to obtaining consistent spectroscopic factors and to show that relativistic corrections have a much greater influence on providing this consistency.

\item[Methods:] This calculation employs the recoil corrected continuum shell model, a model that uses a realistic interaction and produces non-spurious scatterings states that are solutions to the coupled-channel problems.  Pionic and pair contributions to the exchange currents were calculated as developed by Dubach et al. [Nucl. Phys. A \textbf{271}, 279 (1976)].  Relativistic effects are included by use of the direct Pauli reduction.

\item[Results:]  Contributions of the exchange currents are shown to be insufficient to provide consistent spectroscopic factors.  However, the inclusion of relativistic corrections produces spectroscopic factors from the different responses and cross sections which are very similar for both the $p_{1/2}$ and $p_{3/2}$ states and both the $(\left|{\bm q}\right|,\omega)$ = (460 MeV/c, 100 MeV) and  $(\left|{\bm q}\right|,\omega)$ = (570 MeV/c, 172 MeV) data.  The influence of channel coupling is also shown to be significant.  Tests of current conservation show that inclusion of the direct Pauli reduction produces small increases in its violation.

\item[Conclusions:] Results are model dependent.  Meson-exchange current contributions are not sufficient.  Relativistic corrections can be large and lead to more consistent spectroscopic factors. Contributions from channels other than the outgoing channels can be significant.  Response functions which depend on the transverse current are sensitive to the lower component of relativistic wave functions, and hence, would provide a measure of the appropriateness of any relativistic model.
\end{description}
\end{abstract}


\maketitle

\section{Introduction}

Coincidence measurements with the ($e,e^{\prime}p$) reaction in the quasi-elastic region have proven useful in determining the single-particle nature of nuclear states.  Hence, several measurements of $^{16}$O($e,e^{\prime}p$) have been performed in an effort to obtain information on the single-particle wave functions and spectroscopic factors.

By assuming one-photon exchange, current conservation, and zero electron mass, one may approximate the cross section for this reaction as
\begin{eqnarray}
\label{Eq1}
\frac{d^{6} \sigma}{d \Omega_{k^{\prime}} d \omega d \Omega_{p^{\prime}} dE^{\prime}} = \frac{mp^{\prime}}{\left(2 \pi\right)^{3}} \sigma_{Mott} \left[ \nu_{L} R_{L} + \nu_{T} R_{T}\right. \nonumber\\
\left.+ \nu_{LT} R_{LT} \cos{ \alpha} +\nu_{TT} R_{TT} \cos{ 2 \alpha} \right] ,
\end{eqnarray}
where the $R_{i}$ are the response or structure functions, and the $\nu_{i}$ are functions of the kinematic variables.  Early experiments were performed in parallel kinematics (the ejected proton is parallel or anti-parallel to the three-momentum transfer $\bm q$) in which case the interference terms $R_{LT}$ and $R_{TT}$ vanish, allowing one to separate $R_{L}$ and $R_{T}$. Later experiments were performed with $\alpha \neq 0^{\circ}$ or 180$^{\circ}$.  This allowed further separation of the cross section and determination of  $R_{T}$, $R_{LT}$ and $R_{L}-\left({\bm q}^{2}/2Q^{2}\right)R_{TT}$, where $Q^{2}=\left|\omega^{2}-{\bm q}^{2}\right|^{2}$.

Spectroscopic factors are then determined from these measurements by assuming a model in which the $^{15}$N states are pure $p^{-1}_{1/2}\left(p\right)$ or $p^{-1}_{3/2}\left(p\right)$ coupled to an outgoing proton and observing by what fraction the calculated response functions must be reduced to match the experimental values.  One would assume that the same reduction factor should apply to each response function.  However, in Ref. \cite{SBJ93}, which reported on an experiment centered at $(\left|{\bm q}\right|,\omega)$ = (460 MeV/c, 100 MeV),  calculations were performed with a model that gave a proper account of previous data taken in parallel kinematics.  They found that an enhancement factor for $R_{LT}$ for the residual $p^{-1}_{1/2}\left(p\right)$ state differed by a factor of 1.5 and that for the $p^{-1}_{3/2}\left(p\right)$ state differed by more that a factor of 2 from factors required for the other response functions.  Difficulty with this response may not be unexpected since it is an interference response.

In an effort to explain the discrepancy, pion exchange current contributions were calculated in Ref. \cite{SRW94}.  The authors include a seagull term, a pion-in-flight term, and an intermediate $\Delta$ creation term.  The seagull term is their dominant contribution.  These terms were calculated within a continuum random-phase approximation model.  With reduction factors of 0.60 $p^{-1}_{1/2}\left(p\right)$ and 0.51 $p^{-1}_{3/2}\left(p\right)$ applied to the calculated response functions with exchange currents, the discrepancy among responses was mostly removed, although the resulting shapes as a function of missing momentum were not that good.  This was somewhat surprising since the authors of Ref. \cite{BR91} predicted a small contribution from meson-exchange currents (MECs).  The authors of Ref. \cite{SRW94} do point out that different factors of 0.49 $p^{-1}_{1/2}\left(p\right)$ and 0.41 $p^{-1}_{3/2}\left(p\right)$ were required to fit $R_{LT}$  and cross sections for a previous experiment \cite{C91} at $(\left|{\bm q}\right|,\omega)$ = (570 MeV/c, 172 MeV).

Therefore, the problem of different reduction factors appeared to be solved.  But the puzzle continued with the later calculation of Ref. \cite{A99} where the authors address the unexpected success of the MECs.  In this reference the outgoing proton is a solution to a complex optical potential, and the MECs are treated as in Ref. \cite{R79} and, hence, contain seagull, pion-in-flight and isobar graphs. The authors test the sensitivity of the MEC contribution by making calculations with two different optical potentials \cite{Schw82,CK80} and a Woods-Saxon based continuum shell model \cite{A92}. They find considerable sensitivity to the models, but in all cases the total contribution of MECs to $R_{LT}$ is far too small to eliminate the discrepancy between $R_{LT}$  and the other structure functions.  The resulting agreement with the data is unsatisfactory.

It should be noted that the results of Ref. \cite{A99} demonstrated a significant model dependence. Hence, this present article investigates the reaction in a different model, the recoil corrected continuum shell model (RCCSM) \cite{Ph77,Ha05}. The RCCSM has the advantage of producing wave functions that are antisymmetric and contain no spurious components since the calculations are performed in the center-of-mass system.  Orthogonality between bound and scattering states is guaranteed.  The input to the RCCSM consists of only an oscillator-size parameter, $\nu_{0}=m\omega / \hbar$, the desired states of the $A-1$ core nuclei (labeled by $\left|\alpha J_{A}\right\rangle$), and a realistic, translationally invariant interaction.  An oscillator constant of $\nu_{0}$ = 0.3068 fm$^{-2}$ is employed in this paper.  The results show little variation with respect to reasonable $p$-shell values of the constant.  The interaction is taken as that for $^{12}$C from Ref. \cite{Ha13}.  It was derived from a fit to Cohen and Kurath \cite{CoKu65} matrix elements plus $g$-matrix elements of the Reid soft core \cite{R68} for $^{16}$O.  The $p^{-1}_{1/2}\left(p\right)$  and $p^{-1}_{3/2}\left(p\right)$  states of $^{15}$N and the $p^{-1}_{1/2}\left(n\right)$  and $p^{-1}_{3/2}\left(n\right)$ states of $^{15}$O are taken as the core states.  Therefore, one is solving the $^{15}$N($1/2^{-}$) + p, $^{15}$N($3/2^{-}$) + p, $^{15}$O($1/2^{-}$) + n, and $^{15}$O($3/2^{-}$) + n coupled-channels problem.  The choice of pure hole states for core states means that the reduction factors required to fit the data can be converted to spectroscopic factors.  This does not mean that the hole states are oscillator wave functions, but a linear combination of oscillator functions which are solutions to the translationally invariant Hamiltonian.

Although producing considerable improvement in the troublesome $R_{TL}$ from Ref. \cite{SBJ93}, the RCCSM response functions still showed inconsistencies.  Therefore, the next step was to include the MECs.  Pionic (corresponding to the pion-in-flight contribution of Refs. \cite{SRW94,A99}) and pair contributions (corresponding to the seagull term of Refs. \cite{SRW94,A99}) to the exchange currents were calculated with the operators of Ref. \cite{DKD76}.  No isobar contribution is included.  Isobar currents contributions tend to be small and are model dependent. 
The isobar currents have been studied extensively in Refs. \cite{Am03,Ta03}.
 An additional aspect of the procedure in Ref. \cite{DKD76} is the care that is taken to remove contributions from the pair current that would already be contained in the one-pion-exchange part of the two-body potential.  The resulting contribution of MECs to $R_{TL}$ was in the correct direction, but too small, very similar those of Ref. \cite{A99} and not like those in Ref. \cite{SRW94}.

In Refs. \cite{SRW94,C91} the authors indicated that relativity plays a rather unimportant role in quasielastic scattering.  Such a conclusion is understandable when, for example, one is comparing the near perfect agreement of a nonrelativistic optical model calculation \cite{LC94} with results of relativistic calculations \cite{M01,M02} for the cross section data of Ref. \cite{LC94}.  However, this conclusion can be tested for this case by adding relativistic corrections to the RCCSM with a direct Pauli reduction \cite{FPS94,GH13}. This procedure adds to the conventional multipole operators, terms of order $1/M_{N}^{2}$.  The inclusion of these terms increases the cross sections 10$\%$-15$\%$; increases $R_{T}$ and $R_{LT}$, reduces $R_{L}-\left({\bm q}^{2}/2Q^{2}\right)R_{TT}$, and produces near agreement with reduction factors for responses and cross sections for the p$_{1/2}$ state for both the $(\left|{\bm q}\right|,\omega)$  = (460 MeV/c, 100 MeV) and the $(\left|{\bm q}\right|,\omega)$  = (570 MeV/c, 172 MeV) experiments.   It also produces near agreement of reduction factors for responses and cross sections for the p$_{3/2}$ state for the $(\left|{\bm q}\right|,\omega)$  = (460 MeV/c, 100 MeV) and $(\left|{\bm q}\right|,\omega)$  = (570 MeV/c, 172 MeV) experiments, but this required a different reduction factors for the two experiments.

Tests were performed to determine if the direct Pauli reduction terms increased current conservation violation.  The increase was small.  Tests were performed to determine the contribution of components of the final-state wave function which couple to the exit channel.  It was found that these contributions can be significant.  These results put into question the conclusions from one-channel calculations, such as optical models.

\section{The Model}

The exclusive cross section in the laboratory for scattering to a definite residual nuclear state with $\hbar=c=1$ is given by
\begin{eqnarray}
\label{Eq2}
\frac{d^{5} \sigma}{d \Omega_{e}d \Omega_{p}  d \omega } = \frac{\alpha^{2}}{q_{\mu}^{4}} \left( \frac{2k_{0}^{\prime}p_{0}p}{k_{0}R} \right)  \left( k_{\mu} k_{\nu}^{\prime}  +  k_{\mu}^{\prime} k_{\nu} \right. \nonumber\\
\left.+ q_{\mu}^{2} g_{\mu \nu}/2 \right) J^{\mu}J^{\nu *},
\end{eqnarray}
where $\alpha$ is the fine structure constant; the incident and exit electron momenta are $k_{\mu}=(k_{0}, {\bm k})$ and $k_{\mu}^{\prime}=(k_{0}^{\prime}, {\bm k^{\prime}})$; the final, free nucleon momentum is $p_{\mu}=(p_{0}, {\bm p})$; the final core momentum is $p_{A \mu}=(E_{A}, {\bm p_{A}})$; the momentum transferred to the nucleus is $q_{\mu}=(q_{0}, {\bm q})= k_{\mu}^{\prime}-k_{\mu}$ ; and $R$ comes from the density of states and is called the recoil factor,
\begin{equation}
\label{Eq3}
R = \left(1-p_{0}{\bm p} \cdot {\bm p_{A}}/p^{2}E_{A} \right) ,
\end{equation}

Equation (\ref{Eq2}) may be reduced to the form of  Eq. (\ref{Eq1}) by assuming current conservation and zero electron mass.  This procedure is described in Ref. \cite{BG96}.  One can obtain
\begin{eqnarray}
\label{Eq4}
\frac{d^{5} \sigma}{d \Omega_{k^{\prime}} d \omega d \Omega_{p^{\prime}}} = \frac{E_{p}p}{\left(2 \pi\right)^{3} R} \sigma_{Mott} \left[ \nu_{L} R_{L} + \nu_{T} R_{T}\right. \nonumber\\
\left.+ \nu_{LT} R_{LT} \cos{ \alpha} +\nu_{TT} R_{TT} \cos{ 2 \alpha} \right] ,
\end{eqnarray}
where the $\nu_{i}$'s are given in Ref. \cite{SBJ93}, except that $-Q^{2}$ must be substituted for $Q^{2}$ to be consistent with the definition of  $Q^{2}$ in this work and in Ref. \cite{BG96}.

The $R_{i}$'s are given in terms of the currents. The $z$ direction is chosen to be in the ${\bm q}$ direction  and the $y$ direction as in the ${\bm k^{\prime} \times {\bm k}}$ direction.  With this choice of coordinate system, the currents may be written as
\begin{eqnarray}
\label{Eq5}
J_{\lambda} =- \left( 2 \pi \right)^{1/2} \sum_{J \geq 1} \left(-i \right)^{J} \left(2J+1 \right)^{1/2}   \left[ -T_{j \lambda}^{el}+ \lambda T_{J \lambda}^{mag}   \right] \\
{\rm for} \ \  \lambda= \pm 1 , \nonumber
\end{eqnarray}
\begin{equation}
\label{Eq6}
J_{\lambda=0} =J_{z} = \left( 4 \pi \right)^{1/2} \sum_{J = 0} \left(-i \right)^{J} \left(2J+1 \right)^{1/2}  L_{J} ,
\end{equation}
\begin{equation}
\label{Eq7}
J_{0} = \left( 4 \pi \right)^{1/2} \sum_{J = 0} \left(-i \right)^{J} \left(2J+1 \right)^{1/2}  M_{J}^{Coul} ,
\end{equation}
where the multipole operators in Eqs. (\ref{Eq5}) and (\ref{Eq7}) are defined in Eqs. (3.21), (3.32), and (3.33) of Ref. \cite{FW65}.  The operator in Eq. (\ref{Eq6}) is defined below in Eq. (\ref{Eq25}).  Matrix elements of these operators between initial and final states can be multiplied in pairs to produce the $R_{i}$'s, $R_{L}=\left\langle J_{0}J_{0}^{*}\right\rangle$, $R_{T}=\left\langle J_{1}J_{1}^{*}+  J_{-1}J_{-1}^{*}\right\rangle$, $R_{LT}=\left\langle J_{x}J_{0}^{*}+  J_{0}J_{x}^{*}\right\rangle$, and $R_{TT}=\left\langle J_{-1}J_{1}^{*}+  J_{1}J_{-1}^{*}\right\rangle$ where $J_{x}=\left(J_{-1}-J_{1}\right)/\sqrt{2}$ and the angular brackets indicate a sum over the final angular projections of the proton and residual nucleus and an average over the initial nucleus state projections.  Note that $R_{TL}$ is an interference term and, therefore, likely to be sensitive to the model employed.

The wave function with outgoing flux $\nu_{i}$ with initial conditions $i=\left\{\alpha J_{A}M_{A}m_{s}\right\}$ takes the form \cite{HaPh81,Ha96}
\begin{eqnarray}
\label{Eq8}
\psi^{\left(-\right)}_{i} = \left(4 \pi / p_{i}\right) \sum i^{l} Y^{*}_{lm_{l}} \left(\hat{p}\right) \exp(-i\sigma_{l}) \left(-i/2\right) \times \nonumber\\
C^{l \ 1/2 \ j}_{m_{l} m_{s} m} C^{J_{A} \ j\ J_{B}}_{M_{A} m M_{B}} \Psi^{J_{B}M_{B}\left(-\right)}_{c} ,
\end{eqnarray}
where the sum is over $lm_{l}jmJ_{B}M_{B}$ and 
\begin{equation}
\label{Eq9}
\Psi^{J_{B}M_{B}\left(-\right)}_{c} = \sum_{c^{\prime}} r^{-1} u^{J_{B}\left(-\right)}_{c^{\prime}} \left(r\right) \left|\alpha^{\prime} J^{\prime}_{A}l^{\prime}j^{\prime}J_{B}M_{B}\right\rangle .
\end{equation}
The radial function $u^{J_{B}\left(-\right)}_{c^{\prime}}$ has the asymptotic form
\begin{equation}
\label{Eq10}
u^{J_{B}\left(-\right)}_{c^{\prime}} = u^{J_{B}\left(+\right) *}_{c^{\prime}} \rightarrow (\nu_{c} / \nu_{c^{\prime}})^{1/2} \left(O_{c^{\prime}} \delta_{cc^{\prime}}-I_{c^{\prime}}S_{cc^{\prime}}\right) .
\end{equation}
The index $c$ stands for $\alpha J_{A}lj$ with $J_{A}$ and $j$ coupled to $J_{B}$, and $p_{i}$ is the nucleon momentum in the nucleon-nucleus center-of-mass frame.  As a reference point, note that when $\vec{p}_{i}$ is in the $-\hat{z}$ (defined above) direction, $Y^{*}_{lm_{l}}\left(\hat{p_{i}}\right)=\left(-1\right)^{l}\sqrt{2l+1}/4 \pi$. The Hermitian conjugate of this wave function is found by \textbf{R}-matrix techniques \cite{Ph75}. The expansion basis consists of oscillator wave functions for $\bar{n}+\bar{l} \leq 24 = \rho_{max}$, where $\bar{n}$ starts at zero.  These wave functions must be divided by $\left(2 \pi\right)^{3/2}$ for use in Eq. (\ref{Eq2}), but not Eq. (\ref{Eq4}) where the factor is factored out.  The RCCSM involves a transformation to the center-of-mass system.  Therefore, the energy for which the Hamiltonian is solved is the total $p$ + core center-of-mass energy.  The invariants in Eq. (\ref{Eq2}) are calculated in that frame.  More information on the application of the RCCSM to $p$-shell nuclei, may be found in Refs. \cite{Ha13,Ha02}.

The first calculations are performed with the conventional zeroth- and first-order multipoles,
\begin{equation}
\label{Eq11}
M^{Coul}_{JM} = \sum_{i} j_{J}\left(qr_{i}\right) Y_{J M}\left(\hat{r}_{i}\right) F^{i}_{1}\left(q^{2}_{\mu}\right) ,
\end{equation}
{\small\begin{align}
\label{Eq12}
T^{el}_{J M} & = \sum_{i} \left(F^{i}_{1}\left(q^{2}_{\mu}\right) / M_{N}\right) \nonumber\\
& \times  \left\{-\left(\frac{J}{2J+1}\right)^{1/2} j_{J+1}\left(qr_{i}\right) \left[Y_{J+1}\left(\hat{r}_{i}\right)\otimes\vec{\nabla}_{i} \right]_{J M}    \right.\nonumber\\
&+ \left. \left(\frac{J+1}{2J+1}\right)^{1/2} j_{J-1}\left(qr_{i}\right) \left[Y_{J-1}\left(\hat{r}_{i}\right)\otimes\vec{\nabla}_{i} \right]_{J M}     \right\}\nonumber\\
&+\left[F^{i}_{1}\left(q^{2}_{\mu}\right) +K_{i}F^{i}_{2}\left(q^{2}_{\mu}\right) \right] \left[q/\left(2M_{N}\right) \right] j_{J}\left(qr_{i}\right)  \nonumber\\
& \times \left[Y_{J}\left(\hat{r}_{i}\right)\otimes\sigma_{i}\right]_{J M}  ,
\end{align}}
{\small\begin{align}
\label{Eq13}
T^{mag}_{J M} &= \sum_{i} \left(iq\right) \left\{-\left(\frac{J}{2J+1}\right)^{1/2} \right. \nonumber\\
& \times j_{J+1}\left(qr_{i}\right) \left[Y_{J+1}\left(\hat{r}_{i}\right)\otimes \sigma_{i} \right]_{J M}    \nonumber\\
&+\left. \left(\frac{J+1}{2J+1}\right)^{1/2} j_{J-1}\left(qr_{i}\right) \left[Y_{J-1}\left(\hat{r}_{i}\right)\otimes\sigma_{i} \right]_{J M}     \right\}\nonumber\\
& \times \left[F^{i}_{1}\left(q^{2}_{\mu}\right) +K_{i}F^{i}_{2}\left(q^{2}_{\mu}\right) \right] /\left(2M_{N}\right)  \nonumber\\
&-\left[iF^{i}_{1}\left(q^{2}_{\mu}\right)/M_{N} \right] j_{J}\left(qr_{i}\right) \left[Y_{J}\left(\hat{r}_{i}\right)\otimes\vec{\nabla}_{i}\right]_{J M} .
\end{align}}
 With the replacements $F^{i}_{1}\left(q^{2}_{\mu}\right)=F^{i}_{2}\left(q^{2}_{\mu}\right)=1$ or 0 for $p$, $n$ and $F^{i}_{1}\left(q^{2}_{\mu}\right) +K_{i}F^{i}_{2}\left(q^{2}_{\mu}\right)=\mu_{i}$, these expressions become the commonly employed multipoles of Ref. \cite{FW65}. The nucleon form factors employed are those of Ref. \cite{JaHo66}.  

Higher order terms are obtained when one takes the covariant electromagnetic current density,
{\small\begin{equation}
\label{Eq14}
\hat{J}^{\mu}\left(x\right) = e_{i} \bar{\psi}_{f}\left(x\right) \gamma^{\mu} \psi_{i}\left(x\right)+\frac{e_{i}}{2M} \partial_{\mu}\left[ \bar{\psi}_{f}\left(x\right) K \sigma^{\mu\nu}  \psi_{i}\left(x\right)  \right] .
\end{equation}}
However, the multipoles resulting from Eq. (\ref{Eq14}) are matrices and act on wave functions with upper and lower components,
\begin{equation} 
\label{Eq15}
\psi =
\left( \begin{array}{*{20}{c}}
{\left[F\left(r\right)/r\right]\Phi_{\kappa m}}\\
{\left[iG\left(r\right)/r\right]\Phi_{- \kappa m}}
\end{array} \right)
=
\left( \begin{array}{*{20}{c}}
{\eta_{U}}\\
{\eta_{L}}
\end{array} \right)
 ,
\end{equation}
However, one can assume that the upper component of a wave function is given by the non-relativistic calculation and the lower component by the single-particle Dirac equation with no potential , 
\begin{equation}
\label{Eq16}
\eta_{L} \rightarrow \tilde{\eta}_{L}=\frac{\vec{\sigma}\cdot \vec{p}}{E+M}\eta_{U} .
\end{equation}
This approximation with the current of Eq. (\ref{Eq14}) yields Eqs. (\ref{Eq11})-(\ref{Eq13}) plus terms of order $1/M^{2}_{N}$.  Use of Eq. (\ref{Eq16}) has been made by other authors \cite{U95,ODW72}. The approach has been referred to as direct Pauli reduction \cite{FPS94}. It appears that such a substitution introduces more complicated multipole operators, however, it was shown in Ref. \cite{GH13} that Eq. (\ref{Eq16}) is equivalent to substituting $G(r)$ with $\tilde{G} = (F^{\prime} + \kappa F/r)/(E + M)$.  Hence, no new operators need be calculated, and the terms of order $1/M^{2}_{N}$ may be introduced into conventional shell model calculations and conventional electron-scattering codes.

One should note that the multipole operators of Ref. \cite{F85} and Eq. (8)-(10) of Ref. \cite{GH13} are consistent with those of Ref. \cite{FW65}.  However, when employing them in calculations of the nuclear current density, a minus sign appears in front of the transverse electric multipole as shown in Eq. (\ref{Eq5}) above.

As mentioned above, the MECs are calculated in the method of Ref. \cite{DKD76}.  If one performs a nonrelativistic reduction of the diagrams in that work, one arrives at two transverse nuclear currents,
{\small\begin{eqnarray}
\label{Eq17}
j_{pair} = \frac{2g^{2}e_{\pi}}{\left(2 \pi\right)^{3}\left(4M^{2}_{N}\right)} \times \nonumber\\
\frac{\delta^{3}\left(\vec{p}_{1}+\vec{p}_{2}-\vec{p}^{\prime}_{1}-\vec{p}^{\prime}_{2}-\vec{q}\right)\vec{\sigma}_{2} \cdot \left(\vec{p}_{2}-\vec{p}^{\prime}_{2}\right)\vec{\sigma}_{1}}{-\left(E^{\prime}_{2}-E_{2}\right)^{2}+\left(\vec{p}^{\prime}_{2}-\vec{p}_{2}\right)^{2}+\mu^{2}-i\epsilon} - \left(1 \rightarrow 2\right)
\end{eqnarray}}
and
{\small\begin{eqnarray}
\label{Eq18}
j_{pionic} = \frac{2g^{2}e_{\pi}}{\left(2 \pi\right)^{3}\left(4M^{2}_{N}\right)} \times \nonumber\\
\frac{\delta^{3}\left(\vec{p}_{1}+\vec{p}_{2}-\vec{p}^{\prime}_{1}-\vec{p}^{\prime}_{2}-\vec{q}\right)\vec{\sigma}_{1} \cdot \left(\vec{p}_{1}-\vec{p}^{\prime}_{1}\right)}{-\left(E^{\prime}_{1}-E_{1}\right)^{2}+\left(\vec{p}^{\prime}_{1}-\vec{p}_{1}\right)^{2}+\mu^{2}-i\epsilon} \times\nonumber\\
\frac{\vec{\sigma}_{2}  \cdot \left(\vec{p}_{2}-\vec{p}^{\prime}_{2}\right) \left(\vec{p}_{1}-\vec{p}^{\prime}_{1}\right)}{-\left(E^{\prime}_{2}-E_{2}\right)^{2}+\left(\vec{p}^{\prime}_{2}-\vec{p}_{2}\right)^{2}+\mu^{2}-i\epsilon} - \left(1 \rightarrow 2\right)
\end{eqnarray}}
In order to embed this expression in a shell model calculation, one needs to evaluate its expectation value over the momentum distribution of two particles in the nucleus,
{\small\begin{eqnarray}
\label{Eq19}
\left\langle \vec{j} \cdot \hat{\epsilon}\right\rangle = \int d\vec{p}_{1} d\vec{p}_{2} d\vec{p}^{\prime}_{1} d\vec{p}^{\prime}_{2} 
 \left[ \Phi_{j^{\prime}_{1}} \left(\vec{p}^{\prime}_{1}\right) \Phi_{j^{\prime}_{2}}\left(\vec{p}^{\prime}_{2}\right) \right]^{J_{f} \ *}_{M_{f}}  \nonumber\\
\times \left(\vec{j} \cdot \hat{\epsilon}\right) \left[ \Phi_{j_{1}}\left(\vec{p}_{1}\right) \Phi_{j_{2}}\left(\vec{p}_{2}\right) \right]^{J_{i}}_{M_{i}}
\end{eqnarray}}
It is customary \cite{DKD76,ChL70} to Fourier transform the momentum space wave functions to obtain the coordinate space operators given in Ref. \cite{DKD76}.  However, as one can imagine from the condition  $\bar{n}+\bar{l} \leq 24 = \rho_{max}$, the number of required oscillator matrix elements is large, and the machine time to calculate them is prohibitive.  Fortunately, the Fourier transform of an oscillator wave function is an oscillator, and the matrix elements are more efficiently calculated in momentum space.  These momentum space expressions are given in the Appendix.  

\section{Results}

\begin{figure*}[!htbp]
\includegraphics[width=14cm,angle=0]{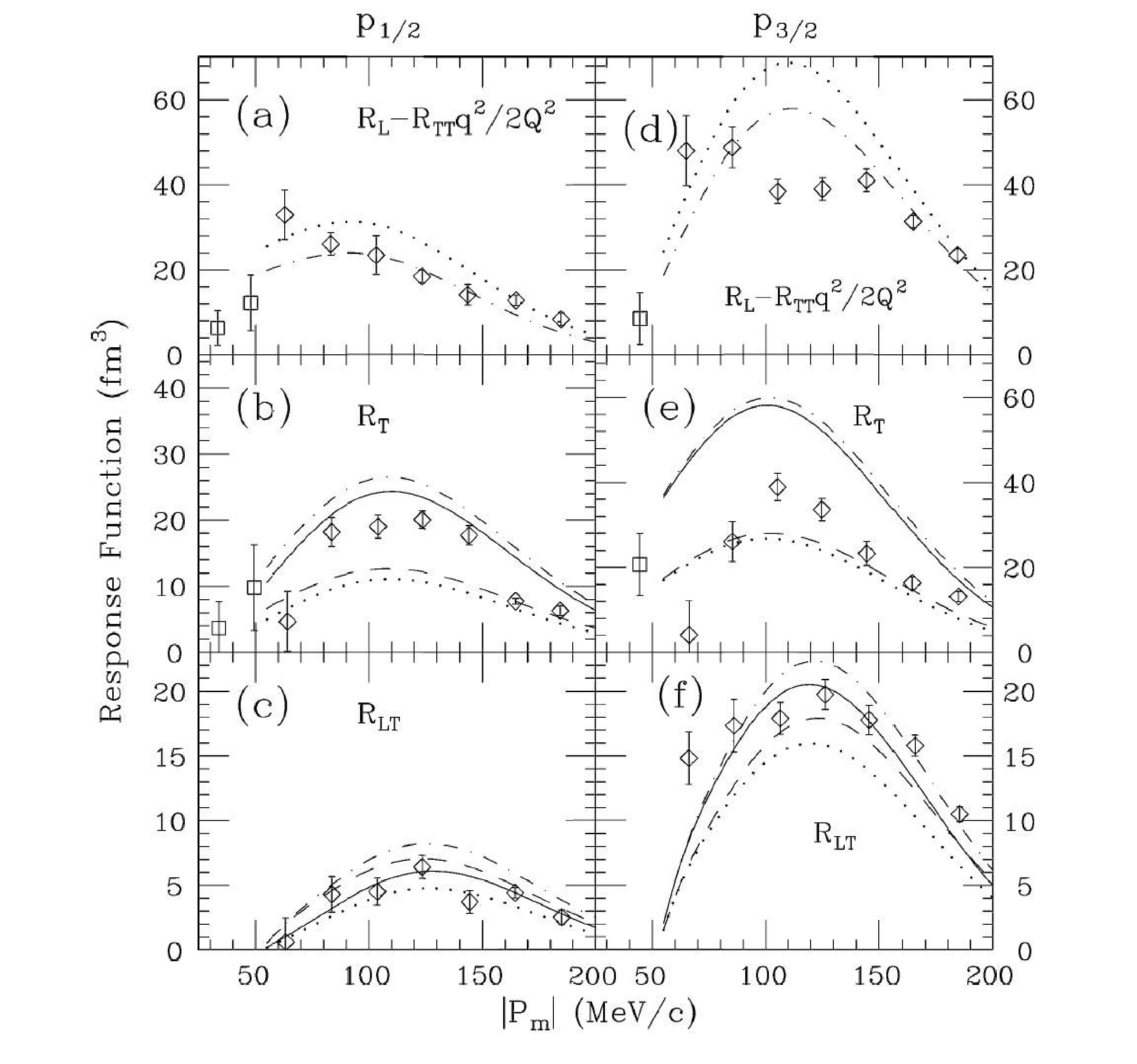}
\caption{\label{Fig1} Response functions as a function of the missing momentum $\left|{\bm P_{m}}\right|$ for $^{16}$O to the $p_{1/2}$ and $p_{3/2}$ states of $^{15}$N.  The dotted lines are from the conventional one-body multipoles; the solid lines include direct Pauli reduction terms.  The dashed lines include MECs. The dot-dashed lines include both direct Pauli reduction terms and MECs.  Data are from Ref. \cite{SBJ93}.  Open squares are from parallel kinematics.}
\end{figure*}
\begin{figure}[!htbp]
\includegraphics[width=9cm,angle=0]{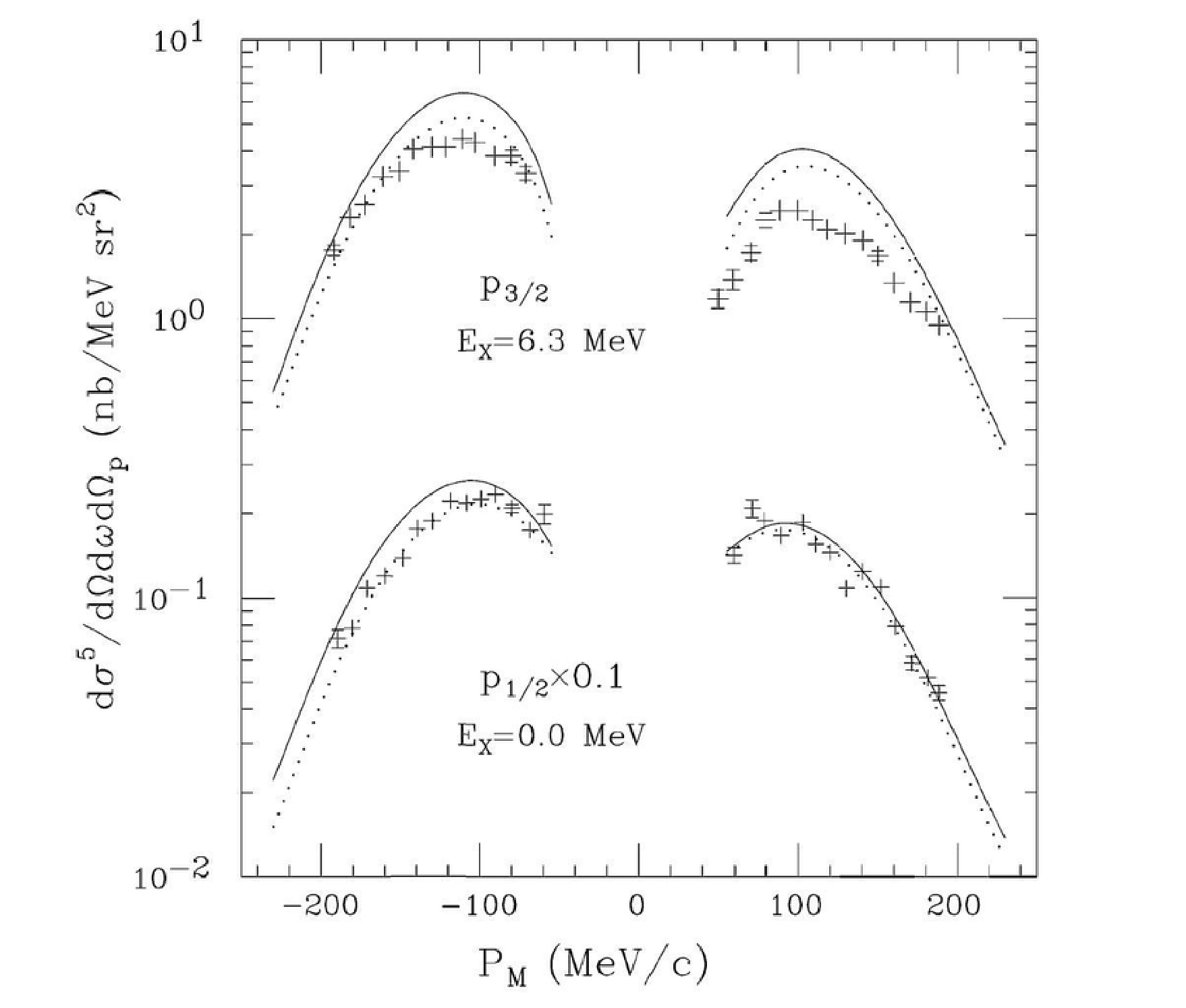}
\caption{\label{Fig2} Cross sections for $^{16}$O to the $p_{1/2}$ and $p_{3/2}$ states of $^{15}$N.  The dotted lines are from the conventional one-body multipoles; the solid lines include direct Pauli reduction terms.  The data are from Ref. \cite{SBJ93} as given in Ref. \cite{SRW94}.}
\end{figure}
\begin{figure}[!htbp]
\includegraphics[width=8cm,angle=0]{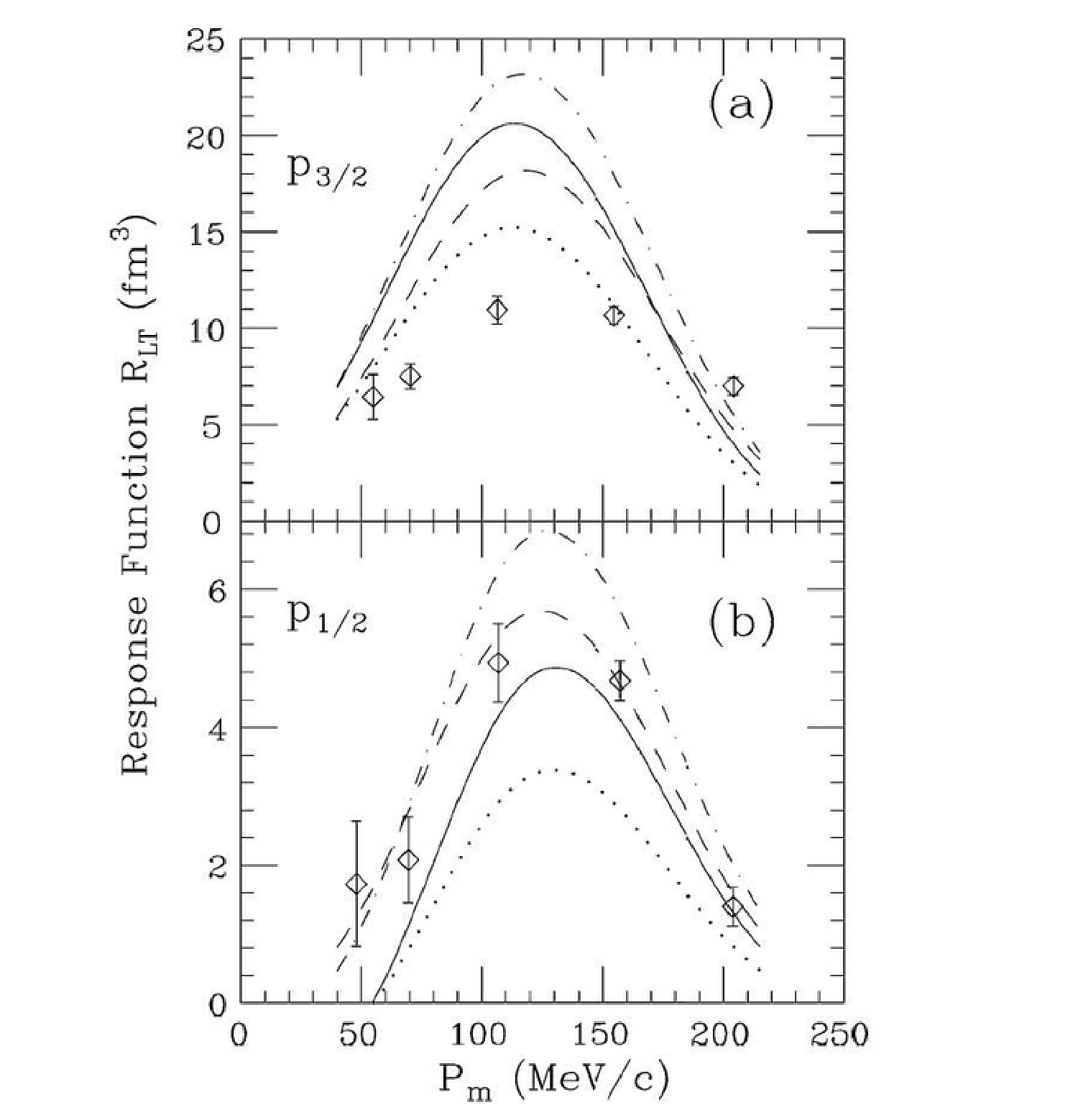}
\caption{\label{Fig3}  Response functions as a function of the missing momentum $\left|{\bm P_{m}}\right|$ for $^{16}$O to the $p_{1/2}$ and $p_{3/2}$ states of $^{15}$N.  The dotted lines are from the conventional one-body multipoles; the solid lines include direct Pauli reduction terms.  The dashed lines include MECs.  The dot-dashed lines include both direct Pauli reduction terms and MECs.  Data are from Ref. \cite{C91}.}
\end{figure}
\begin{figure}[!htbp]
\includegraphics[width=8cm,angle=0]{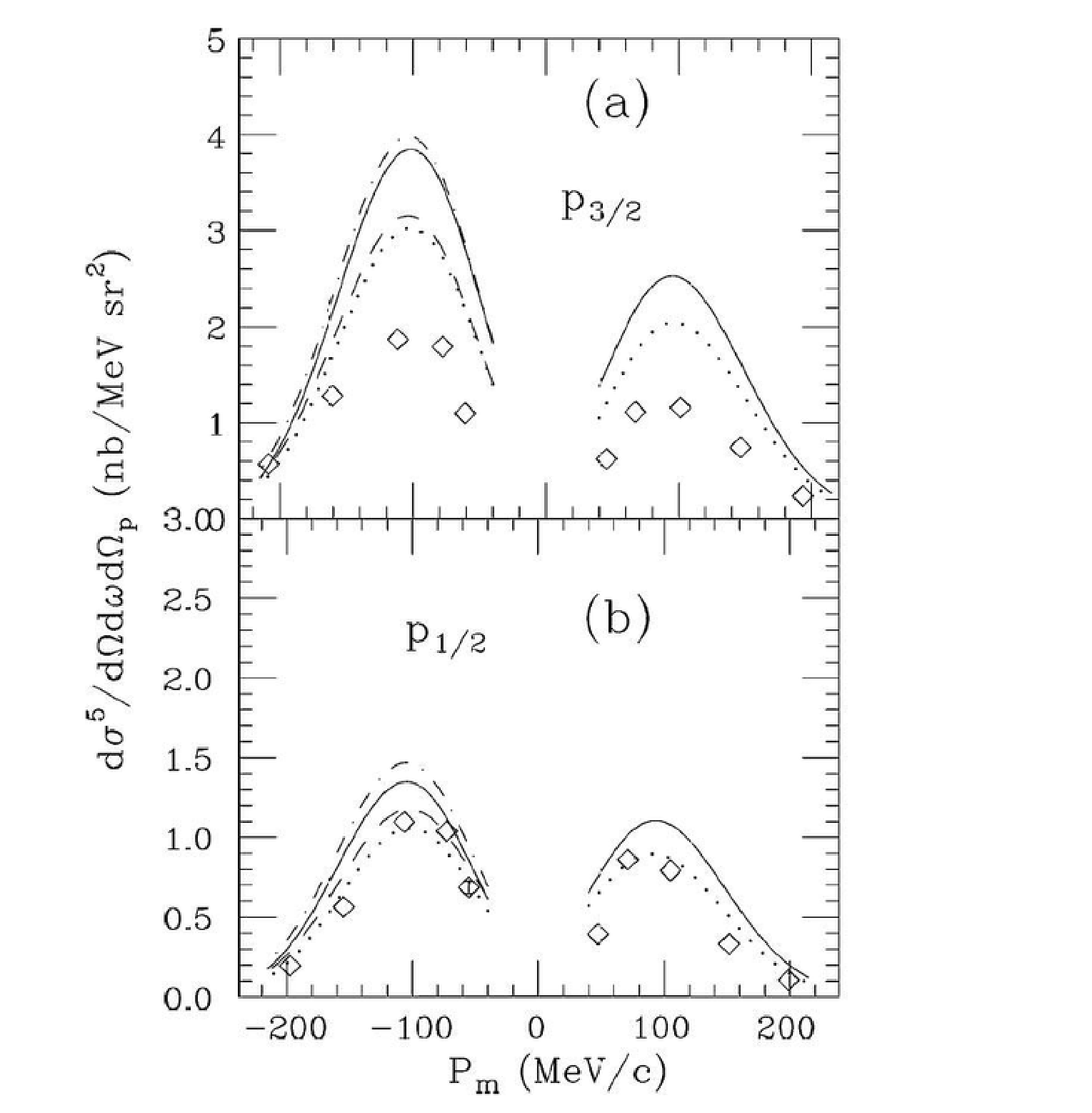}
\caption{\label{Fig4} Cross sections for $^{16}$O to the $p_{1/2}$ and $p_{3/2}$ states of $^{15}$N.  The dotted lines are from the conventional one-body multipoles; the solid lines include direct Pauli reduction terms.  The data are from Ref. \cite{C91}.}
\end{figure}
In Fig. \ref{Fig1} are shown the calculated response functions and the data of Ref. \cite{SBJ93}.  The calculated response functions are those defined above $\times$ $E^{\prime}_{p}/\left(M_{N}R\right)$ so that Eq. (\ref{Eq4}) is compatible with Eq. (\ref{Eq1}) used in the data analysis.  This factor is close to unity. In Fig. \ref{Fig2} are plotted the cross sections from these data.  These data are centered at $(\left|{\bm q}\right|,\omega)$ = (460 MeV/c, 100 MeV), a range where the RCCSM has been successful in describing particle-induced reactions.  The dotted lines in all panels of Fig. \ref{Fig1} correspond to the calculations with the conventional, one-body multipoles in Eqs. (\ref{Eq11})-(\ref{Eq13}).  The solid lines include the direct Pauli reduction.  The dashed lines include MECs, and the dot-dashed lines include both.  In Fig. \ref{Fig2} and panels (a) and (d) of Fig. \ref{Fig1} the lines with only MECs are omitted because they would be nearly indistinguishable from those without MECs.  Figure \ref{Fig3} shows the same calculations for $R_{LT}$, at $(\left|{\bm q}\right|,\omega)$ = (580 MeV/c, 172 MeV), with the data of Ref. \cite{C91}, and Fig. \ref{Fig4} shows the cross sections for this reference.  The authors of Ref. \cite{C91} are to be commended for publishing cross sections instead of some model-dependent extraction from the cross sections.  In Fig. \ref{Fig4} only the cross section without MECs are plotted for $P_{m} > 0$, since those with MECs would be indistinguishable from those without MECs.  One notices that the conventional calculations (dotted lines) for the $p_{1/2}$ state cross sections in Fig. \ref{Fig2} and \ref{Fig4} agree very well with the data of Refs. \cite{SBJ93,C91}, but not the responses in Fig. \ref{Fig1} and \ref{Fig3}.  In Fig. \ref{Fig1} the dotted curves fall below that data for $R_{LT}$, woefully below the data for $R_{T}$, and above the data for $R_{L}-\left({\bm q}^{2}/2Q^{2}\right)R_{TT}$.  Similarly, for the $p_{1/2}$ state in Fig. \ref{Fig3}, $R_{LT}$ falls below.

Now consider the dashed lines showing the contribution of the MECs to the $p_{1/2}$ state at $(\left|{\bm q}\right|,\omega)$ = (460 MeV/c, 100 MeV).  Their addition to the cross section in Fig. \ref{Fig2} is negligible, and they contribute only a few percent to the responses.  These results are very similar to those obtained in Fig. 2 of Ref. \cite{A99}.  At $(\left|{\bm q}\right|,\omega)$ = (580 MeV/c, 172 MeV), the contribution to the cross section in Fig. \ref{Fig4} is again very small, however, $R_{LT}$ receives a considerable boost at this higher momentum transfer.

In the calculation of the MEC matrix elements, some approximations are made.  Similar approximations are made for the one-body current matrix elements.  First, the $^{16}$O ground state is composed of the closed oscillator shell, taken as the vacuum $\left|0\right\rangle$ plus excitations of the form $\left[np_{1/2} \otimes 0 p^{-1}_{1/2}\right]^{0}$ and $\left[np_{3/2} \otimes 0 p^{-1}_{3/2}\right]^{0}$.  The matrix elements of the transition operators that connect  $\left|0\right\rangle$ to the scattering states are correctly transformed to the center of mass.  This is referred to as being recoil corrected.  The closed shell comprises 70.5$\%$ of the ground state.  The matrix elements connecting the other ground-state components are not recoil corrected in order to save calculation time.  Such a correction should be small for $^{16}$O.  Second, the terms corresponding to the recoil of nucleons in the core are neglected as described in Ref. \cite{Ha13}.  These were included in $^{4}$He$\left(e,e^{\prime}p\right)^{3}$H by transforming to Jacobi coordinates \cite{Ha96}. Although important in $^{4}$He, these should be small for $^{16}$O.  The only correction associated with the Jacobi coordinates is to use ${\bm r}=\left(A-1\right)/A \bm{\xi}$ in the multipole equations, where $\bm\xi$ is the coordinate connecting the outgoing proton to the center of mass of the core.  Hence, \textbf{r} is measured from the center of mass of the whole system.  Third, matrix elements of particle-hole states, $\left\langle  j_{p}m_{t_{p}}j^{-1}_{h}m_{t_{h}}\left(J\right)   \left\|T^{J}\right\|  j^{\prime}_{p}m_{t_{p^{\prime}}}j^{-1}_{h^{\prime}}m_{t_{h^{\prime}}}\left(0\right)   \right\rangle$, contain terms of the form
{\small\begin{eqnarray}
\label{Eq20}
- \delta_{n_{p}l_{p}j_{p}m_{t_{p}},n_{p^{\prime}}l_{p^{\prime}}j_{p^{\prime}}m_{t_{p^{\prime}}}} \sum_{n_{c}l_{c}j_{c}m_{t_{c}}}  \frac{\hat{J_{x}}\hat{J^{\prime}}}{\hat{j_{p}}} W\left(J_{x}Jj_{c}j_{h};J^{\prime}_{x}j_{p}\right)  \nonumber\\
\times \left\langle j_{c}m_{t_{p}}j_{h^{\prime}}m_{t_{h^{\prime}}}\left(J_{x}\right) \left\| T^{J} \left[ \right. \right\|  j_{c}m_{t_{c}}j_{h}m_{t_{h}}\left(J^{\prime}_{x}\right) \right\rangle \nonumber\\
-(-1)^{j_{h}+j_{c}-J^{\prime}_{x}} \left. \left. \left. \right\|  j_{h}m_{t_{h}} j_{c}m_{t_{c}}\left(J^{\prime}_{x}\right) \right\rangle \right]
\end{eqnarray}}
Such terms are ignored in for both one- and two-body transition operators since it is unlikely that a scattering particle state would be the same as the bound particle state. 

Now consider adding the direct Pauli reduction terms to the conventional calculations shown as the solid lines.  All of the responses that depend on the transverse current are increased, and $R_{T}$ in Figs. \ref{Fig1}(b) and \ref{Fig1}(e) is increased by a factor of 2.  $R_{L}-\left({\bm q}^{2}/2Q^{2}\right)R_{TT}$ is decreased due to a reduction in $R_{L}$.  It has been a pattern in this paper that the direct Pauli reduction increases the transverse response and decreases the longitudinal response. In Ref. \cite{Am03} the model employed produced relativistic corrections that were small and in the opposite direction for $R_{T}$ and $R_{L}$.
  It was pointed out previously, and confirmed in this case, that the transverse response is sensitive to the lower components of a relativistic calculation.  At this point one would say that the calculation for the $p_{1/2}$ state is in reasonable agreement with the data.

\begin{figure*}[!htbp]
\includegraphics[width=14cm,angle=0]{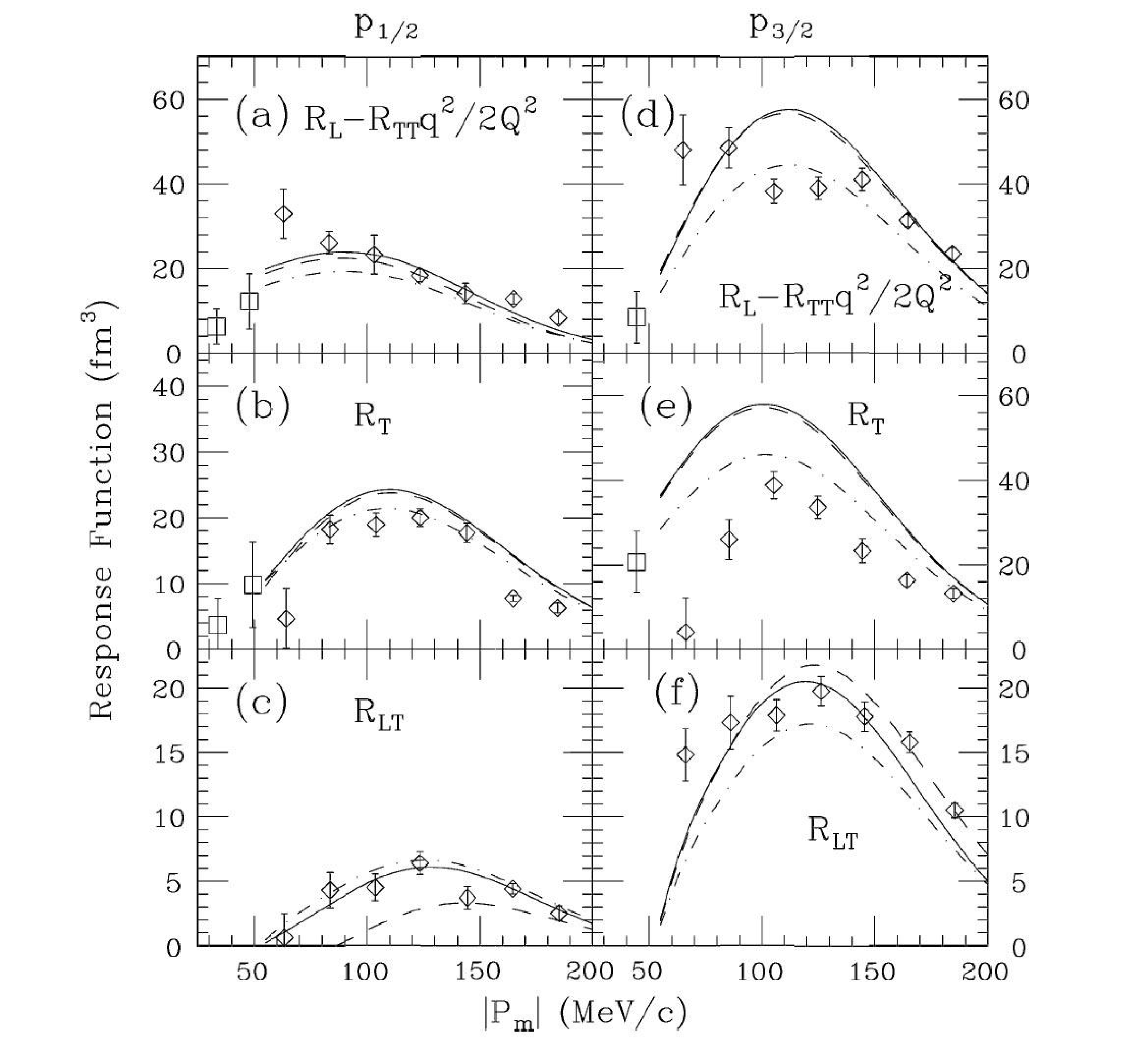}
\caption{\label{Fig5} Response functions as a function of the missing momentum $\left|{\bm P_{m}}\right|$ for $^{16}$O to the $p_{1/2}$ and $p_{3/2}$ states of $^{15}$N.  The solid lines include direct Pauli reduction terms.  The dashed lines are the same except they are from omitting channels other than the exit channel.  The dot-dashed lines for the $p_{1/2}$ state show the direct Pauli reduction plus MEC calculations $\times$ 0.81.  The dot-dashed lines for the $p_{3/2}$ state show the direct Pauli reduction plus MEC calculations $\times$ 0.77.  Data are from Ref. \cite{SBJ93}.   Open squares are from parallel kinematics.}
\end{figure*}

The inclusions of both the direct Pauli reduction and MEC are shown as dot-dashed lines in Figs. \ref{Fig1}, \ref{Fig3}, and \ref{Fig4}.  The results for the $p_{1/2}$ state are reasonably consistent.  With a reductions factor of 0.81 one obtains very good fits to $R_{LT}$ and the cross section at $(\left|{\bm q}\right|,\omega)$ = (570 MeV/c, 172 MeV) and to $R_{T}$ , $R_{LT}$ , and the cross sections at $(\left|{\bm q}\right|,\omega)$ = (460 MeV/c, 100 MeV) .  This agreement is demonstrated in panels (b) and (c) of Fig. \ref{Fig5}.  Only the $R_{L}-\left({\bm q}^{2}/2Q^{2}\right)R_{TT}$  at $(\left|{\bm q}\right|,\omega)$  = (460 MeV/c, 100 MeV) reduced response lies slightly below the data as shown as the dot-dashes line in  panel (a) of Fig. \ref{Fig5}.  This is in contrast to the variations seen in Table I of Ref. \cite{SBJ93}, where the complete distorted-wave impulse approximation results are below the data.  The resulting spectroscopic factor of 1.62 is larger that that obtained in Ref. \cite{LC94} from cross section data.

Now consider the difficult $p_{3/2}$ state shown in Figs. \ref{Fig1}(d)-\ref{Fig1}(f) with the $(\left|{\bm q}\right|,\omega)$ = (460 MeV/c, 100 MeV)  data of Ref. \cite{SBJ93}.  The conventional calculation (dotted line) for $R_{L}-\left({\bm q}^{2}/2Q^{2}\right)R_{TT}$  would require a reduction factor of 0.75 whereas $R_{T}$ and $R_{TL}$ would require enhancement factors of 1.35 and 1.33, respectively.  This result for $R_{TL}$ is actually better than obtained in Refs. \cite{SBJ93,SRW94}.  In fact, the calculated $R_{TL}$ does not stand out as being blatantly in disagreement with the data as it was in those references.  This shows again the model dependence of these calculations.  However, to obtain some consistency among the responses, $R_{L}-\left({\bm q}^{2}/2Q^{2}\right)R_{TT}$  must be lowered and/or $R_{T}$ and $R_{TL}$ must be increased.  Adding MECs does provide some increase in $R_{T}$ and $R_{TL}$, but it is small, and negligible for $R_{L}-\left({\bm q}^{2}/2Q^{2}\right)R_{TT}$.  However, the inclusion of direct Pauli reduction terms increases $R_{T}$ and $R_{TL}$ and decreases $R_{L}-\left({\bm q}^{2}/2Q^{2}\right)R_{TT}$.  That is, everything moves in the correct direction.  The dot-dashed line shows the result of adding both MECs and direct Pauli reduction terms.  All cross sections and responses for the $p_{3/2}$ state are too large.  Figures \ref{Fig5}(d) and \ref{Fig5}(e) show a best fit to $R_{L}-\left({\bm q}^{2}/2Q^{2}\right)R_{TT}$ and $R_{T}$ and the resulting $R_{TL}$ for that reduction factor of 0.77 in \ref{Fig5}(f).  The fitted curve for $R_{L}-\left({\bm q}^{2}/2Q^{2}\right)R_{TT}$ falls somewhat below the data, and the fitted curve for $R_{T}$ falls somewhat above the data.  The fitted curve for $R_{TL}$ falls somewhat below the data, but not the factors of 2 and greater found in other calculations with no relativistic corrections \cite{SBJ93,C91,A99}.

The $p_{3/2}$  state cross section and $R_{LT}$ at $(\left|{\bm q}\right|,\omega)$ = (570 MeV/c, 172 MeV) are well fitted by a reduction factor of 0.56.  In Ref. \cite{SRW94} it was necessary to have different reduction factors for the data of Refs. \cite{SBJ93,C91} for both the $p_{1/2}$  and $p_{3/2}$  calculations.  Requiring different reduction factors at different momenta transferred may not be surprising for this $p_{3/2}$  state.  This state obviously has a large component of higher-order configurations which may have a different momentum transfer dependence that the pure hole state. 

Finally, Fig. \ref{Fig5} shows the result of omitting channels other than the exit channel.  The solid lines include the direct Pauli reduction, and the dashed lines show the same calculation omitting channels other than the exit channel.  This means the sum over $c^{\prime}$ in Eq. (\ref{Eq9}) is limited to only $c^{\prime}=c$.  Optical models are generally used for the outgoing proton.  Optical models reduce the flux of the proton in its exit channel by absorption.  In a coupled-channel calculation, flux is reduced in the exit channel by sending it to other channels.  However, what one sees in Fig. \ref{Fig5} is that these channels contribute to the currents and in this case show up strongly in $R_{TL}$ for the $p_{1/2}$ state.  These contributions are even more important at lower momenta transferred \cite{Ha13}. This demonstrates a difficulty with optical model calculations of knockout reactions, whether nonrelativistic or relativistic.

\section{Current conservation violation}

Although the direct Pauli reduction procedure has produced terms that simulate the effect of the lower component of a relativistic model, the question arises as to whether the new terms produce significant current conservation violation.  Serious current conservation violation would make inclusion of the terms less credible.  Therefore, a test is made for the magnitude of current conservation violation.

The nuclear current operator can be written in momentum space as 
\begin{equation}
\label{Eq21}
J^{\mu}\left(\vec{q}\right)=\int \exp\left(-i \vec{q} \cdot \vec{x}\right) \left\langle f \left| \hat{J}\left(\vec{x}\right)\right| i \right\rangle d\vec{x},
\end{equation}
the four-vector operator $J^{\mu}\left(\vec{q}\right)$ having components $\left(\rho,\vec{J}\right)$.  A coordinate system is chosen where $\hat{q}=\hat{z}$ and $\hat{x}=\hat{\theta}_{q}$, and spherical unit vectors defined as in Ref. \cite{FW65}: $\hat{\epsilon}_{q0}=\hat{\epsilon}_{z}=\hat{z}$ , and $\hat{\epsilon}_{q\pm 1}= \mp \left(\hat{\epsilon}_{x} \pm \hat{\epsilon_{y}}\right)/\sqrt{2}$.  Since the magnetic quantum numbers of the nucleus and emitted nucleon will eventually be summed over, the choice of coordinate system is arbitrary.  The investigation of current conservation begins by examining the longitudinal current, $J_{0}\left(\vec{q}\right)=\hat{\epsilon}_{q0} \cdot \vec{J}\left(\vec{q}\right)$.  One performs the expansion, 
{\small\begin{eqnarray}
\label{Eq22}
 \hat{\epsilon}_{q0} \exp\left(-i \vec{q} \cdot \vec{x}\right)= \hat{\epsilon}_{q0} \exp\left(-i q z\right)= \nonumber\\
\sum_{J} \left(-i\right)^{J} i \sqrt{4 \pi \left(2J+1\right)} \left( \sqrt{\frac{J+1}{2J+1}} j_{J+1}\left(qx\right)\vec{Y}^{0}_{J+1,J}\left(\hat{x}\right) \right.\nonumber\\
+ \left. \sqrt{\frac{J}{2J+1}} j_{J-1}\left(qx\right)\vec{Y}^{0}_{J-1,J}\left(\hat{x}\right)  \right)= \nonumber\\
\left(i/q\right) \sum_{J} \left(-i\right)^{J} \sqrt{4 \pi \left(2J+1\right)} \vec{\nabla} \left[ j_{J}\left(qx\right)Y_{J0}\left(\hat{x}\right)  \right] .
\end{eqnarray}}

With the non-relativistic current operator (the magnetization current does not contribute) one has
{\small\begin{eqnarray}
\label{Eq23}
\hat{\epsilon}_{q0} \cdot \vec{J}\left(\vec{q}\right)= \frac{i}{q} \sum_{i} \frac{e^{i}_{N}}{2mi} \sum_{J} \left(-i\right)^{J} \sqrt{4 \pi \left(2J+1\right)} \times \nonumber\\
\int \vec{\nabla}_{i} \left[ j_{J}\left(qx_{i}\right)Y_{J0}\left(\hat{x}_{i}\right)  \right] \left[ \Psi^{+}_{f} \vec{\nabla}_{i} \Psi_{i} - \left(\vec{\nabla}_{i} \Psi^{+}_{f} \right)\Psi_{i}\right] d\vec{x}_{i}.
\end{eqnarray}}
With the replacement of the charge by the corresponding nucleon form factor, the longitudinal current operator thus becomes
{\small\begin{eqnarray}
\label{Eq24}
J_{0}\left(\vec{q}\right)= \hat{\epsilon}_{q0} \cdot \vec{J}\left(\vec{q}\right)=J_{z}\left(\vec{q}\right)= \sum_{i} \left(-i\right)^{J}  \sqrt{4 \pi \left(2J+1\right)} \times \nonumber\\
\frac{F^{i}_{1}}{M_{N}} \left[ \frac{1}{q} \vec{\nabla}_{i} \left[ j_{J}\left(qx_{i}\right)Y_{J0}\left(\hat{x}_{i}\right)  \right] \cdot \vec{\nabla}_{i} -\frac{q}{2} j_{J}\left(qx_{i}\right)Y_{J0}\left(\hat{x}_{i}\right) \right].
\end{eqnarray}}
The expression for the longitudinal current operator, Eq. (\ref{Eq24}), has single-particle operators to be placed between initial and final wave functions.  The continuity equation $q_{\mu}J^{\mu}\left(\vec{q}\right)=0$ takes the form $qJ_{z}\left(\vec{q}\right)=q_{0}\rho\left(\vec{q}\right)$ with the above choice of coordinate system.  The question now becomes how will this result be used?  One could ignore the continuity equation and just calculate the longitudinal current with the above equation and use that result in the cross section calculation.  Or one could choose not to calculate the longitudinal current and replace it with $q_{0}\rho\left(\vec{q}\right)$ in the spirit of Siegert’s theorem \cite{S37}, assuming one has a better knowledge of the nuclear density than the current.  Alternatively, one could replace $\rho\left(\vec{q}\right)$ with $qJ_{z}\left(\vec{q}\right)/q_{0}$.  It was shown in Ref. \cite{PNK96} that the two choices correspond to two different choices of gauge for the virtual photon propagator.  In fact Ref. \cite{PNK96} investigates three different gauges and demonstrates that, for non-conserved currents, different gauges produce different cross sections.  One can build current conservation into the longitudinal current operator; or is it more honest to calculate it directly and say this is what the model produces?  Some incite into this question is given in Ref. \cite{K95}, where calculations for bound-state excitations are performed with different choices for the current and in limited and extended model spaces.  The authors found that different choices for the current produced different cross sections, but the differences were much smaller when the model space was extended.  In addition, one transverse electric operator with current conservation built into it \cite{FH85}, gave similar results to both the limited and extended model spaces.  This result provides some evidence that invoking current conservation in some form is preferable to calculating the longitudinal current operator matrix elements.  Therefore, the above cross-sectional calculations included replacement of the longitudinal current with $q_{0}\rho\left(\vec{q}\right)$.

One can test the degree of current conservation violation by comparing matrix elements of $q_{0}\rho\left(\vec{q}\right)$ with those of $J_{z}\left(\vec{q}\right)$.  However, the current will then depend upon the residual nucleus, its angular momentum projection, the spin projection of the outgoing nucleon, and its direction.  Comparison with all of these variables would be difficult.  Therefore, the choice is made to look at $J^{*}_{0}\left(\vec{q}\right)J_{0}\left(\vec{q}\right)$, and then choose the ground state of the residual nucleus with charge $Z-1$, sum over magnetic quantum numbers of the residual nucleus and outgoing proton, and integrate over the outgoing proton direction.  One is left with comparing the ratio of the quantity $\sum_{Jlj}\left(2J+1\right)\left| \left\langle \psi^{\left(-\right)}_{B} \left\| L_{J} \right\| 0 \right\rangle \right|^{2}$ to $\sum_{Jlj}\left(2J+1\right)\left| \left\langle \psi^{\left(-\right)}_{B} \left\| q_{0}M^{Coul}_{J} /q \right\| 0 \right\rangle \right|^{2}$, where
{\small\begin{equation}
\label{Eq25}
 L_{J}=\sum_{i} \frac{F^{i}_{1}}{M_{N}} \left[ \frac{1}{q} \vec{\nabla}_{i} \left[ j_{J}\left(qx_{i}\right)Y_{J0}\left(\hat{x}_{i}\right)  \right] \cdot \vec{\nabla}_{i} -\frac{q}{2} j_{J}\left(qx_{i}\right)Y_{J0}\left(\hat{x}_{i}\right) \right] .
\end{equation}}
The comparison to be made is this ratio, the test of current conservation, with and without the direct Pauli reduction terms.  That is, does the attempt to include relativistic effects render current more or less conserved?

\begin{figure*}[!htbp]
\includegraphics[width=14cm,angle=0]{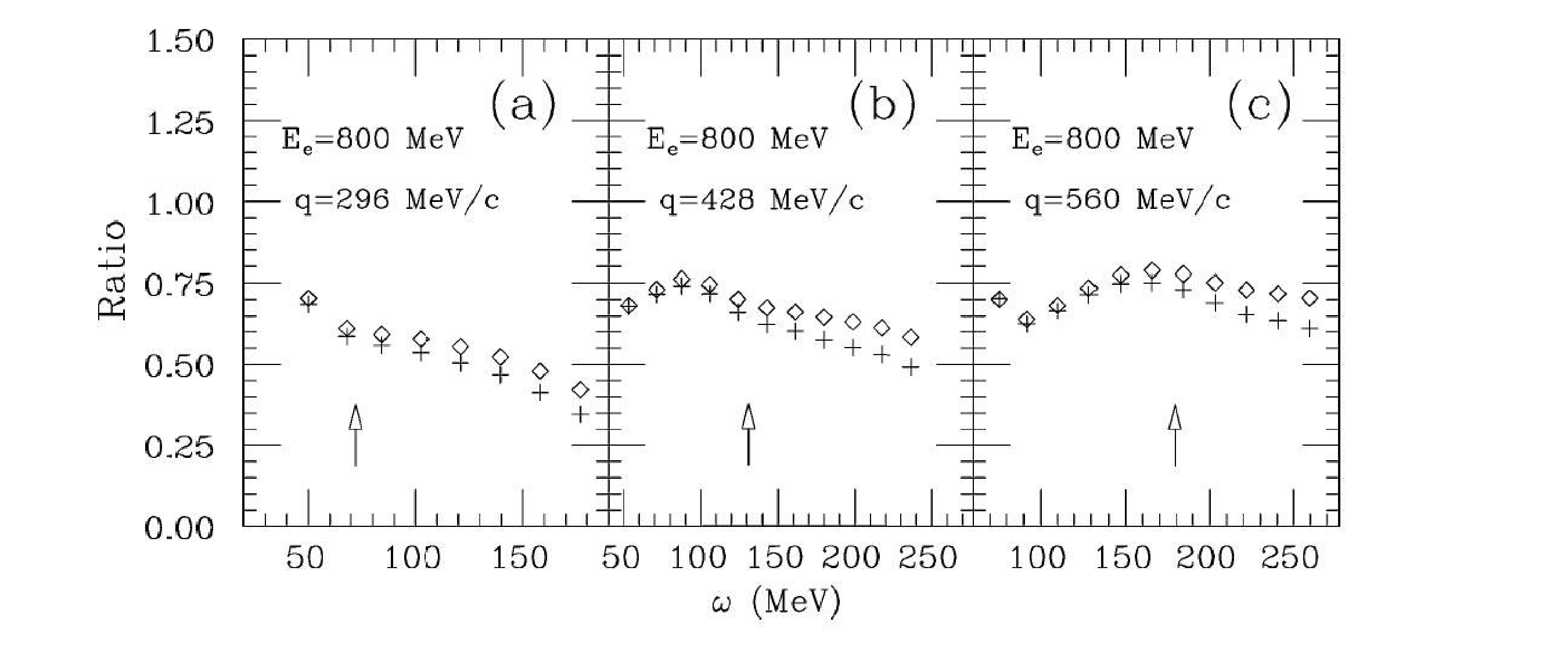}
\caption{\label{Fig6} Ratio of the summed matrix elements of $L_{J}$ to those of $q_{0}M^{Coul}_{J}$.  The open diamonds (crosses) are without (with) direct Pauli reduction terms.  The arrows indicate the position of the quasielastic peak.}
\end{figure*}
Figure \ref{Fig6} shows this ratio for three momenta transferred and energies transferred over the range of the quasielastic region.  The open diamonds correspond to calculations without the direct Pauli reduction terms; the crosses correspond to those with the terms.  The arrows indicate the approximate position of the quasielastic peak.  A slightly smaller ratio at the quasielastic peak, indicating slightly more current conservation violation is found with the inclusion.  This is likely due to the fact that the artificially constructed wave function is further from the exact solution to the model Hamiltonian.  However, since the differences are small near the quasielastic peak, the use of the direct Pauli reduction does not have a serious effect on current conservation. 

\section{Conclusion}

Calculations were performed for $^{16}$O$\left(e,e^{\prime}p\right)^{15}$N$\left(1/2^{-},3/2^{-}\right)$ with the RCCSM to determine the effect of MECs and relativistic corrections on cross sections and response functions.  Of special concern was the inability of optical model calculations to reproduce the interference response $R_{TL}$.  Two previous calculations came to opposite conclusions as to whether MECs could substantially increase $R_{TL}$ and hence, provide some consistency among spectroscopic factors extracted from individual cross sections and response functions.  Inclusion of MECs of Ref. \cite{DKD76} in the RCCSM produced results very much like those of Ref. \cite{A99} and led to the same conclusion - that MECs do not make a substantial contribution to producing consistency.  

Therefore, the effect of relativistic corrections was tested by including direct Pauli reduction terms in the RCCSM.  These terms contributed substantially to the response functions and provided reasonable consistency for spectroscopic factors. A reduction factor of 0.81, applied to all $p_{1/2}$ calculations, at $(\left|{\bm q}\right|,\omega)$  = (460 MeV/c, 100 MeV), and $(\left|{\bm q}\right|,\omega)$  = (570 MeV/c, 172 MeV) provided a good fit to all cross sections and response functions.  Calculations for the p3/2 showed similar consistency for the data at $(\left|{\bm q}\right|,\omega)$  = (460 MeV/c, 100 MeV) data with a reduction factor of 0.77 but required a reduction factor of 0.56 to provide consistency between $R_{TL}$ and the cross section for the $(\left|{\bm q}\right|,\omega)$  = (570 MeV/c, 172 MeV) data.

A test of current conservation showed that inclusion of direct Pauli reduction terms increased the violation only slightly and is, therefore, an appropriate procedure.  The importance of a coupled-channel model was demonstrated by performing calculations that eliminate contributions from channels other than the exit channel.  Therefore, one would think that the ideal calculation for comparing with data would be a coupled-channel calculation in a completely relativistic model.  Such models exist, but would be plagued by uncertainties in the center-of-mass transformation.  

\appendix*
\section{}

The following substitutions are made in Eq. (\ref{Eq18}), ${\bm t}_{1}={\bm p}_{1}-{\bm p}^{\prime}_{1}$, ${\bm T}_{1}=\left({\bm p}_{1}+{\bm p}^{\prime}_{1}\right)/2$, ${\bm t}_{2}={\bm p}_{2}-{\bm p}^{\prime}_{2}$, ${\bm T}_{2}=\left({\bm p}_{2}+{\bm p}^{\prime}_{2}\right)/2$. The delta-function in Eq. (18) is expanded as 
\begin{equation}
\label{EqA1}
\delta^{3} \left({\bm t}_{1} + {\bm t}_{2} -{\bm q}  \right)= \left( 2 \pi \right)^{-3} \int d{\bm p} \exp\left[i \left( {\bm t}_{1} + {\bm t}_{2} -{\bm q}  \right) \cdot {\bm p} \right] . 
\end{equation}
With the choice of $\hat{q}=\hat{z}$ the exponentials in Eq. (\ref{EqA1}) may be expanded as 
{\small\begin{eqnarray}
\label{EqA2}
\exp\left[i \left( {\bm t}_{1} + {\bm t}_{2} -{\bm q}  \right) \cdot {\bm p} \right]= \left( 4 \pi \right)^{3} \sum_{lml^{\prime}m^{\prime}D}  i^{l+l^{\prime}} \left( -i \right)^{D}  j_{l } \left( {\bm t}_{1} \rho \right) \nonumber\\
 \times j_{l^{\prime} } \left( {\bm t}_{2} \rho \right) j_{D } \left( {\bm q} \rho \right) Y_{lm} \left( \hat{t}_{1} \right) Y^{*}_{lm} \left( \hat{\rho} \right)  Y_{l^{\prime}m^{\prime}} \left( \hat{t}_{2} \right) Y_{l^{\prime}m^{\prime}} \left( \hat{\rho} \right) \times \nonumber\\
 Y_{D0} \left( \hat{\rho} \right) \hat{D} /   \left( 4 \pi \right)^{1/2}  ,
\end{eqnarray}}
where $\hat{D}= \left( 2D+1 \right)^{1/2}$ .  The current operator is now separable in the coordinates of particles 1 and 2.  This means that it is not necessary to make the $jj$ to $ls$ coupling transformation.

By coupling together all terms involving particle 1 and all terms involving particle 2, one obtains the reduced matrix elements of the multipole operators in a proton-neutron basis,
{\small\begin{eqnarray}
\label{EqA3}
\left\langle j^{\prime}_{1} j^{\prime}_{2}  \left( J_{f} \right) \left\| T^{mag}_{J \left( \rm pionic\right)}  \left( q \right)  \right\|  j_{1} j_{2}  \left( J_{i} \right)   \right\rangle = \frac{g^{2}}{4 \pi} \frac{Q_{\pi}}{4M^{2}_{N} \pi^{7/2}}  \times \nonumber\\
 \sum_{ll^{\prime}J_{x}LF} i^{a-1}  \left( -1 \right)^{b+F} \hat{J} \hat{J}_{f}  \hat{J}_{i} \hat{l}^{\prime} \left[ l \right] C^{L1F}_{000} C^{l^{\prime}1J_{x}}_{000} C^{l1L}_{000} \times \nonumber\\
\left[ \begin{array}{*{20}{ccc}}
 {j^{\prime}_{1}} & {j^{\prime}_{2} } & {J_{f} } \\
 {j_{1} } & {j_{2}} & {J_{i} } \\
 {L } & {l^{\prime}} & {J }  
\end{array} \right] 
C^{ll^{\prime}J}_{000}
\left[ \begin{array}{*{20}{ccc}}
 {J} & {1} & {J} \\
 {l} & {l^{\prime}} & {L } 
\end{array} \right] 
\int \rho^{2} d \rho  \times \nonumber\\
 I_{2} \left( j^{\prime}_{1}, F, L, l, j_{1} \right)  I_{1} \left( j^{\prime}_{2}, J_{x}, l^{\prime}, j_{2} \right) j_{J} \left( q \rho \right)  -  \left( -1 \right)^{c}  \left( 1 \rightarrow 2 \right) ,
\end{eqnarray}}
and 
{\small\begin{eqnarray}
\label{EqA4}
\left\langle j^{\prime}_{1} j^{\prime}_{2}  \left( J_{f} \right) \left\| T^{el}_{J \left( \rm pionic\right)}  \left( q \right)  \right\|  j_{1} j_{2}  \left( J_{i} \right)   \right\rangle = \frac{g^{2}}{4 \pi} \frac{Q_{\pi}}{4M^{2}_{N} \pi^{7/2}}  \times \nonumber\\
 \sum_{ll^{\prime}J_{x}LF} i^{a+1}  \left( -1 \right)^{b+F} \hat{J}_{f}  \hat{J}_{i} \hat{l}^{\prime} \left[ l \right] C^{L1F}_{000} C^{l^{\prime}1J_{x}}_{000} C^{l1L}_{000} \times \nonumber\\
\left[ \begin{array}{*{20}{ccc}}
 {j^{\prime}_{1}} & {j^{\prime}_{2} } & {J_{f} } \\
 {j_{1} } & {j_{2}} & {J_{i} } \\
 {L } & {l^{\prime}} & {J }  
\end{array} \right] 
\int \rho^{2} d \rho I_{2} \left( j^{\prime}_{1}, F, L, l, j_{1} \right)  I_{1} \left( j^{\prime}_{2}, J_{x}, l^{\prime}, j_{2} \right) \nonumber\\
 \times \left( C^{ll^{\prime}J-1}_{000}  \sqrt{J+1}  
\left[ \begin{array}{*{20}{ccc}}
 {J} & {1} & {J-1} \\
 {l} & {l^{\prime}} & {L } 
\end{array} \right] 
 j_{J-1} \left( q \rho \right) - \right. \nonumber\\
 \left. C^{ll^{\prime}J+1}_{000} \sqrt{J}
 \left[ \begin{array}{*{20}{ccc}}
 {J} & {1} & {J+1} \\
 {l} & {l^{\prime}} & {L } 
\end{array} \right]
 j_{J+1} \left( q \rho \right) \right)
  -  \left( -1 \right)^{c}  \left( 1 \rightarrow 2 \right) ,
\end{eqnarray}}
where $g^{2}/4 \pi=14$.  Similarly, the pair current multipole reduced matrix elements become
{\small\begin{eqnarray}
\label{EqA5}
\left\langle j^{\prime}_{1} j^{\prime}_{2}  \left( J_{f} \right) \left\| T^{mag}_{J \left( \rm pair\right)}  \left( q \right)  \right\|  j_{1} j_{2}  \left( J_{i} \right)   \right\rangle =- \frac{g^{2}}{4 \pi} \frac{Q_{\pi}}{4M^{2}_{N} \pi^{7/2}}  \times \nonumber\\
 \sum_{ll^{\prime}J_{x}L} i^{a-1}  \left( -1 \right)^{b+L} \hat{J} \hat{J}_{f}  \hat{J}_{i} \hat{l}^{\prime} \hat{l} \hat{L}  C^{l^{\prime}1J_{x}}_{000} C^{ll^{\prime}J}_{000} 
\left[ \begin{array}{*{20}{ccc}}
 {j^{\prime}_{1}} & {j^{\prime}_{2} } & {J_{f} } \\
 {j_{1} } & {j_{2}} & {J_{i} } \\
 {L } & {l^{\prime}} & {J }  
\end{array} \right] 
 \nonumber\\
 \times
\left[ \begin{array}{*{20}{ccc}}
 {J} & {1} & {J} \\
 {l} & {l^{\prime}} & {L } 
\end{array} \right] 
\int \rho^{2} d \rho   I_{0} \left( j^{\prime}_{1}, l, L, j_{1} \right)  I_{1} \left( j^{\prime}_{2}, J_{x}, l^{\prime}, j_{2} \right)  \nonumber\\
\times  j_{J} \left( q \rho \right)  -  \left( -1 \right)^{c}  \left( 1 \rightarrow 2 \right) ,
\end{eqnarray}}
and
{\small\begin{eqnarray}
\label{EqA6}
\left\langle j^{\prime}_{1} j^{\prime}_{2}  \left( J_{f} \right) \left\| T^{el}_{J \left( \rm pair\right)}  \left( q \right)  \right\|  j_{1} j_{2}  \left( J_{i} \right)   \right\rangle = -\frac{g^{2}}{4 \pi} \frac{Q_{\pi}}{4M^{2}_{N} \pi^{7/2}}  \times \nonumber\\
 \sum_{ll^{\prime}J_{x}L} i^{a+1}  \left( -1 \right)^{b+L} \hat{J}_{f}  \hat{J}_{i} \hat{l}^{\prime} \hat{l} \hat{L} C^{l^{\prime}1J_{x}}_{000} 
\left[ \begin{array}{*{20}{ccc}}
 {j^{\prime}_{1}} & {j^{\prime}_{2} } & {J_{f} } \\
 {j_{1} } & {j_{2}} & {J_{i} } \\
 {L } & {l^{\prime}} & {J }  
\end{array} \right] 
  \nonumber\\
  \times \int \rho^{2} d \rho I_{0} \left( j^{\prime}_{1}, l, L, j_{1} \right)  I_{1} \left( j^{\prime}_{2}, J_{x}, l^{\prime}, j_{2} \right) \nonumber\\
 \times \left( C^{ll^{\prime}J-1}_{000}  \sqrt{J+1}  
\left[ \begin{array}{*{20}{ccc}}
 {J} & {1} & {J-1} \\
 {l} & {l^{\prime}} & {L } 
\end{array} \right] 
 j_{J-1} \left( q \rho \right) - \right. \nonumber\\
 \left. C^{ll^{\prime}J+1}_{000} \sqrt{J}
 \left[ \begin{array}{*{20}{ccc}}
 {J} & {1} & {J+1} \\
 {l} & {l^{\prime}} & {L } 
\end{array} \right]
 j_{J+1} \left( q \rho \right) \right)
  -  \left( -1 \right)^{c}  \left( 1 \rightarrow 2 \right) ,
\end{eqnarray}}
In the above expressions $\left[l \right]=2l+1 $, the bracketed arrays are $6j$'s and $9j$'s, $C^{j_{1}j_{2}J}_{m_{1}m_{2}M} $ is a Clebsch-Gordon coefficient, and $Q_{\pi}$ is now equal to +1 for $j_{1}\left( p \right)j_{2}\left( n \right) \rightarrow j^{\prime}_{1}\left( n \right)j^{\prime}_{2}\left( p \right)$  and $Q_{\pi}$ equals -1 for $j_{1}\left( n \right)j_{2}\left( p \right) \rightarrow j^{\prime}_{1}\left( p \right)j^{\prime}_{2}\left( n \right)$.  The phases are 
$a=l+l^{\prime}+l_{1}+l_{2}+l^{\prime}_{1}+l^{\prime}_{2}$, $b=l_{1}+l_{2}+n_{1}+n_{2}+ n^{\prime}_{1}+n^{\prime}_{2} +J+J_{x}$ and $c=j_{1}+j_{2}+j^{\prime}_{1}+j^{\prime}_{2}-J_{f}-J_{i}$.  The phase convention is chosen such that if both the momentum space and coordinate space oscillator wave functions are taken to be real and positive at the origin, then the multipole reduced matrix elements are identical as calculated in either space.  This required an additional phase because
\begin{eqnarray}
\phi \left( 1/\alpha , {\bm k} \right) = \left( 2 \pi \right)^{-3/2} \int d{\bm r} \psi \left( \alpha ,{\bm r} \right) e^{ -i{\bm k} \cdot {\bm r} }=\nonumber\\
\left( -i \right)^{l} \left( -1 \right)^{n} \psi \left( 1/\alpha , {\bm k} \right), \nonumber
\end{eqnarray}
where n starts at zero.

The $I$-functions in the above equations are the single-particle reduced matrix elements.
{\small\begin{eqnarray}
\label{EqA7}
I_{0} \left( j^{\prime}_{1}, l, L, j_{1} \right)=
\left\langle j^{\prime}_{1}  \left\|  j_{l} \left( t_{1} \rho \right)  \left[ Y_{l} \left( \hat{t}_{1} \right) \otimes \sigma_{1} \right]^{L}   \right\|  j_{1}   \right\rangle = \nonumber\\
2 \sqrt{6 \pi} \hat{j}^{\prime}_{1} \hat{j}_{1} \hat{l} \hat{L}
\left[ \begin{array}{*{20}{ccc}}
 {l^{\prime}_{1}} & {1/2 } & {j^{\prime}_{1} } \\
 {l_{1} } & {1/2} & {j_{1}  } \\
 {l} & {1} & {L }  
\end{array} \right]
\left( -1 \right)^{l_{1}} \sum_{nN} \left\langle n_{1} l_{1} n^{\prime}_{1} l^{\prime}_{1} l \left. \right| nlN0 \right\rangle  \nonumber\\
\times \int^{\infty}_{0} R^{\sqrt{2}/\alpha}_{N0} \left( T_{1} \right)  T^{2}_{1} d T_{1}   \int^{\infty}_{0} j_{l} \left( t_{1} \rho \right) R^{1/ \sqrt{2}\alpha}_{nl} \left( t_{1} \right)  t^{2}_{1} d t_{1} .
\end{eqnarray}}
{\small\begin{eqnarray}
\label{EqA8}
I_{1} \left( j^{\prime}_{2}, J_{x}, l^{\prime}, j_{2} \right)=  \nonumber\\
\left\langle j^{\prime}_{2}  \left\|  j_{l^{\prime}} \left( t_{2} \rho \right) t_{2} / \left( t^{2}_{2}  + \mu^{2} \right) \left[ Y_{J_{x}} \left( \hat{t}_{2} \right) \otimes \sigma_{2} \right]^{l^{\prime}}   \right\|  j_{2}   \right\rangle =\nonumber\\
2 \sqrt{6 \pi} \hat{j}^{\prime}_{2} \hat{j}_{2} \hat{l}^{\prime} \hat{J}_{x}
\left[ \begin{array}{*{20}{ccc}}
 {l^{\prime}_{2}} & {1/2 } & {j^{\prime}_{2} } \\
 {l_{2} } & {1/2} & {j_{2}  } \\
 {J_{x}} & {1} & {l^{\prime}}  
\end{array} \right]
\left( -1 \right)^{l_{2}} \sum_{nN} \left\langle n_{2} l_{2} n^{\prime}_{2} l^{\prime}_{2} l \left. \right| nJ_{x}N0 \right\rangle \times \nonumber\\
 \int^{\infty}_{0} R^{\sqrt{2}/\alpha}_{N0} \left( T_{2} \right)  T^{2}_{2} d T_{2}   \int^{\infty}_{0} j_{l^{\prime}} \left( t_{2} \rho \right) \frac{t_{2}}{ t^{2}_{2}  + \mu^{2} }  R^{1/ \sqrt{2}\alpha}_{nJ_{x}} \left( t_{2} \right)  t^{2}_{2} d t_{2} .
\end{eqnarray}}
{\small\begin{eqnarray}
\label{EqA9}
I_{2} \left( j^{\prime}_{1}, F, L, l, j_{1} \right) =  \nonumber\\
\left\langle j^{\prime}_{1}  \left\|  j_{l} \left( t_{1} \rho \right) t_{1} / \left( t^{2}_{1}  + \mu^{2} \right) \left[ Y_{F} \left( \hat{t}_{1} \right) \otimes \sigma_{1} \right]^{L}   \right\|  j_{1}   \right\rangle =\nonumber\\
2 \sqrt{6 \pi} \hat{j}^{\prime}_{1} \hat{j}_{1} \hat{F} \hat{L}
\left[ \begin{array}{*{20}{ccc}}
 {l^{\prime}_{1}} & {1/2 } & {j^{\prime}_{1} } \\
 {l_{1} } & {1/2} & {j_{1}  } \\
 {F} & {1} & {L}  
\end{array} \right]
\left( -1 \right)^{l_{1}} \sum_{nN} \left\langle n_{1} l_{1} n^{\prime}_{1} l^{\prime}_{1} F \left. \right| nFN0 \right\rangle \times \nonumber\\
 \int^{\infty}_{0} R^{\sqrt{2}/\alpha}_{N0} \left( T_{1} \right)  T^{2}_{1} d T_{1}   \int^{\infty}_{0} j_{l} \left( t_{1} \rho \right) \frac{t_{1}}{ t^{2}_{1}  + \mu^{2} }  R^{1/ \sqrt{2}\alpha}_{nF} \left( t_{1} \right)  t^{2}_{1} d t_{1} .
\end{eqnarray}}
The oscillator wave functions are $\psi \left( \alpha, \vec{r} \right)=R^{\alpha}_{nl} \left( r \right) Y_{lm} \left( \hat{r} \right) $ and the coupling is $l1/2\left( j\right) $.  
The calculation, therefore, does require calculation of Brody-Moshinsky-Talmi brackets \cite{BrMo67}, 
$\left\langle n_{1} l_{1} n_{2} l_{2} L \left. \right| nlNL \right\rangle$, but only a very restricted few.  
An analytical expression exists for the relative integral in Eq. (\ref{EqA7}), $ \int^{\infty}_{0} j_{l} \left( t_{1} \rho \right) R^{1/ \sqrt{2}\alpha}_{nl} \left( t_{1} \right)  t^{2}_{1} d t_{1} = \left( \pi /2\right)^{1/2} \left( -1 \right)^{n} R^{ \sqrt{2}\alpha}_{nl} \left( \rho \right)$.
The electric multipoles in Eqs. (\ref{EqA4}) and (\ref{EqA6}) must be multiplied by $-1$ if one wants them to be consistent with Refs. \cite{GH13,FW65,F85} and Eq. (\ref{Eq12}) above!

\end{document}